\documentclass[journal]{IEEEtran}

\usepackage[english]{babel}
\usepackage[utf8x]{inputenc}
\usepackage[T1]{fontenc}

\usepackage{amsmath,amssymb,mathrsfs,bbm}
\usepackage{graphicx}
\usepackage[colorinlistoftodos]{todonotes}
\usepackage[colorlinks=true, allcolors=blue]{hyperref}
\usepackage{multirow}
\usepackage{algorithm}
\usepackage{algorithmic}

\usepackage{psfrag,epsfig,graphics}
\usepackage{amsmath,amsthm,amssymb,multirow}
\usepackage{mathbbol}
\usepackage{amssymb}   
\usepackage{url}  
\usepackage{pgfplotstable}

\usepackage{mathabx} 

\DeclareSymbolFontAlphabet{\amsmathbb}{AMSb}%

\newcommand{\cp}[1]{\ifmmode {\mathcal{#1}}\else ${\mathcal{#1}}$\fi}
	
\newcommand{\bA}{\boldsymbol{A}}

\newcommand{\bD}{\boldsymbol{D}}

\newcommand{\bE}{\boldsymbol{E}}
\newcommand{\bF}{\boldsymbol{F}}
\newcommand{\bG}{\boldsymbol{G}}

\newcommand{\bI}{\boldsymbol{I}}

\newcommand{\bM}{\boldsymbol{M}}

\newcommand{\bQ}{\boldsymbol{Q}}
\newcommand{\bR}{\boldsymbol{R}}

\newcommand{\bV}{\boldsymbol{V}}
\newcommand{\bW}{\boldsymbol{W}}
\newcommand{\bX}{\boldsymbol{X}}
\newcommand{\bY}{\boldsymbol{Y}}
\newcommand{\bZ}{\boldsymbol{Z}}

\newcommand{\bZhat}{\,\widehat{\bZ}}

\newcommand{\calD}{\mathcal{D}}

\newcommand{\calG}{\mathcal{G}}

\newcommand{\calM}{\mathcal{M}}
\newcommand{\calN}{\mathcal{N}}

\newcommand{\bDelta}{\boldsymbol{\Delta}}

\newcommand{\bPsi}{\boldsymbol{\Psi}}
\newcommand{\bPhi}{\boldsymbol{\Phi}}
\newcommand{\bSigma}{\boldsymbol{\Sigma}}

\newcommand{\bUpsilon}{\boldsymbol{\Upsilon}}

\newcommand{\cb}[1]{\boldsymbol{#1}}

\newcommand{\vect}{\operatorname{vec}}
\newcommand{\diag}{\operatorname{diag}}

\usepackage{color}  
\def\cred{\textcolor{black}}

\def\cmag{\textcolor{black}}
\definecolor{darkgreen}{rgb}{0., 0.4, 0.}

\title{Deep Hyperspectral and Multispectral Image \\ Fusion with Inter-image Variability}

\author{Xiuheng~Wang,~\IEEEmembership{Student Member,~IEEE,}
        Ricardo~Augusto~Borsoi,~\IEEEmembership{Member,~IEEE,}\\
         C\'edric~Richard,~\IEEEmembership{Senior Member,~IEEE,}
        and Jie~Chen,~\IEEEmembership{Senior Member,~IEEE.}
\thanks{
Xiuheng Wang and C\'edric Richard are with Universit\'{e} C\^{o}te d'Azur, CNRS, OCA, F-06108, Nice, France (e-mail: xiuheng.wang@oca.eu, cedric.richard@unice.fr).  Ricardo Augusto Borsoi is with Universit\'e de Lorraine, CNRS, CRAN, F-54000, Nancy, France (e-mail: raborsoi@gmail.com).
Jie Chen is with Centre of Intelligent Acoustics and Immersive Communications at School of Marine Science and Technology, Northwestern Polytechnical University, Xi'an, 710072, China (e-mail: dr.jie.chen@ieee.org).}
}

\begin{document}
\maketitle

\begin{abstract}
Hyperspectral and multispectral image fusion allows us to overcome the hardware limitations of hyperspectral imaging systems inherent to their lower spatial resolution. Nevertheless, existing algorithms usually fail to consider realistic image acquisition conditions. This paper presents a general imaging model that considers inter-image variability of data from heterogeneous sources and flexible image priors. The fusion problem is stated as an optimization problem in the maximum a posteriori framework. We introduce an original image fusion method that, on the one hand, solves the optimization problem accounting for inter-image variability with an iteratively reweighted scheme and, on the other hand, that leverages lightweight CNN-based networks to learn realistic image priors from data. In addition,  we propose a zero-shot strategy to directly learn the image-specific prior of the latent images in an unsupervised manner. The performance of the algorithm is illustrated with real data subject to inter-image variability.

\end{abstract}


\section{Introduction}
\label{sec:intro}
Hyperspectral imaging systems acquire scenes by recording hundreds of narrow, contiguous spectral bands ranging from visible up to infrared wavelengths. Their rich spectral information has attracted interest in many applications such as remote sensing for mineral exploration, vegetation monitoring, and land cover analysis~\cite{bioucas2013hyperspectral}. Nevertheless, the high spectral resolution of hyperspectral images (HIs) limits their spatial resolution because of hardware limitations~\cite{shaw2003spectralImagRemote}. In contrast, multispectral cameras can achieve a much higher spatial resolution but over a small number of spectral bands. {Consequently}, a strategy to improve the spatial resolution of HIs is to fuse them with multispectral images (MIs) of the same scene. This results in the hyperspectral and multispectral image fusion (HMIF) problem. 

Several strategies have been proposed to solve the HMIF problem. These strategies can be roughly divided into component substitution or multiresolution analysis methods, matrix or tensor factorization methods, and deep learning approaches. Component substitution or multiresolution analysis methods aim to substitute some patterns of the HI, high-frequency ones in particular, by information extracted from the MI~\cite{yokoya2012coupledNMF,liu2000MRA_HS_MS_fusion,aiazzi2006GLP_HS}. These techniques employ different representations of the images, e.g., in the wavelet domain, which are also used for pansharpening~\cite{vivone2018injectionCoefficientsPansharpening,loncan2015pansharpeningReview}.

Subspace-based formulations have become very popular to address HMIF problems since they significantly reduce their dimensionality~\cite{yokoya2012coupledNMF,simoes2015HySure}. They also have a close connection with the widely used linear mixing model~\cite{Keshava:2002p5667, dobigeon2013nonlinear}, which represents each pixel of an HI as a linear combination of a small number of spectral signatures. Several subspace-based formulations have been proposed, often employing prior information about the basis vectors or their contributions in the decomposition, to improve the results. Examples include sparse dictionary learning~\cite{wei2015hs_ms_fusionBayesianSparse,akhtar2015hs_ms_fusionBayesianSparse} or matrix factorization~\cite{yokoya2012coupledNMF} approaches, which can use, e.g., spatial~\cite{simoes2015HySure} and sparse~\cite{kawakami2011hs_ms_fusionMFsparse,lanaras2015hs_ms_fusionMFprojectedGrad} regularizers or patch-level processing~\cite{veganzones2016hs_ms_fusionMFlocalLowRank}. Efficient algorithms also convert this problem into solving a Sylvester equation~\cite{wei2015hs_ms_fusionSylvesterEq}.
Some approaches have considered the manifold structure of the image patches~\cite{zhang2018exploiting}. Other approaches have explored the representation of HIs and MIs as three dimensional tensors~\cite{kanatsoulis2018hyperspectralSRR_coupledCPD,li2018hs_ms_fusionTensorFactorization,prevost2020coupledTucker_hyperspectralSRR_TSP}. Low-rank tensor models have been used to represent the high-resolution images (HRIs), such as the canonical polyadic decomposition~\cite{kanatsoulis2018hyperspectralSRR_coupledCPD}, the Tucker decomposition~\cite{li2018hs_ms_fusionTensorFactorization,prevost2020coupledTucker_hyperspectralSRR_TSP,borsoi2020tensorHSRvariability}, and the block term decomposition~\cite{ding2020HSR_interpretable_BTD}. 

{Deep learning approaches have recently become very popular for HMIF~\cite{li2022deepLearningMultimodalFusionReview, yao2020cross, zhang2020unsupervised}.} These approaches leverage the capability of neural networks to represent complex signals and images. Early supervised approaches were based upon classical neural network architectures used in image processing such as 3D convolutional neural networks (CNN)~\cite{palsson2017HSR_3D_CNN}, while more recent methods explore physical acquisition models to design architectures with improved interpretability~\cite{chen2022integration}, e.g., incorporating CNN results as priors in model-based frameworks~\cite{dian2020regularizing,wang2021hyperspectral} or using architectures inspired by unrolling principle~\cite{xie2020MHFnet_interpretable_HSR}.
However, the scarcity of training data with ground truth has motivated the development of unsupervised approaches, that depend only on the observed HI and MI. Examples include the use of autoencoders with shared weights~\cite{qu2018unsupervisedDusionDirichletNet,liu2022modelInspiredAutoencoder_HSR,wang2020unsupervisedVariational_HSR}, and approaches based on deep image priors~\cite{ulyanov2018deep}, which parameterize the HRI as the output of a neural network and train the latter using different options for the network inputs~\cite{zhang2020HSR_deep_imPrior,wei2020unsupervised}.


Although different strategies have been investigated to solve the HMIF problem, these methods assume that the observed HI and MI are acquired at the same time instant and under the same conditions. However, platforms carrying both hyperspectral and multispectral imaging systems are still limited~\cite{Borsoi_2018_Fusion}. On the contrary, due to the wider availability of satellites with multispectral sensors, e.g., the Sentinel, Landsat and Quickbird missions, it has become of great interest to fuse HIs and MIs acquired at different time instants by different instruments~\cite{yokoya2017HS_MS_fusinoRev}. 
When applied in these realistic conditions, most existing methods suffer from severe limitations as they ignore variability between the HI and MI. Inter-image variability includes localized spatial and spectral changes and can occur due to differences in acquisition conditions caused by, e.g., atmospheric, illumination or seasonal variations~\cite{borsoi2020variabilityReview}, as well as abrupt changes~\cite{liu2019reviewCD_GRSM}.

To tackle this issue, several HMIF frameworks addressing inter-image variability have been recently proposed~\cite{Borsoi_2018_Fusion,borsoi2020tensorHSRvariability,prevost2022LL1_HSR,borsoi2021tensorHSRasilomar, brezini2021fusionIntraImageVariability,camacho2022HS_MS_fusion_spatial_variability, fu2021fusionInterImageChangesL21norm}. A detailed review of these methods is provided in Section~\ref{sec:model_var}. These methods formulate the HMIF problem with a key difference when compared to the original approaches: the HI and the MI are assumed to be generated from distinct HRIs, which are allowed to be different because of spatially homogeneous variations~\cite{Borsoi_2018_Fusion,prevost2022LL1_HSR} or spatially localized ones~\cite{borsoi2020tensorHSRvariability}. However, considering inter-image variability renders the HMIF problem significantly more ill-posed, which makes the use of appropriate prior information about the HRIs very important in order to achieve good performance.

Existing HMIF works that consider inter-image variability rely on handcrafted priors, such as low-rank matrix~\cite{Borsoi_2018_Fusion} or tensor~\cite{borsoi2020tensorHSRvariability,prevost2022LL1_HSR} decompositions. However, these priors are not adequate to model complex contents embedded in real HIs. Without considering inter-image variability, this issue has been addressed in the HMIF problem by exploring the powerful representation capability of deep learning methods, as noted by various recent works on this topic. Nevertheless, devising learning-based approaches to address inter-image variability in HMIF incurs additional challenges, first because very little data is available for training. Indeed, since inter-image variability originates from complex physical phenomena, it is difficult to generate realistic synthetic data to be used for training even if HIs of a single scene are available. This makes learning an end-to-end mapping from an HI and an MI to the HRIs unfeasible.

Recently, deep image priors~\cite{ulyanov2018deep} and plug-and-play strategies~\cite{venkatakrishnan2013plug} have been used to introduce prior information with either pre-trained or unsupervised neural networks.
However, adequately addressing inter-image variability requires considering two different HRIs, underlying the HI and the MI, respectively.
Thus, directly exploiting such strategies to address inter-image variability in HMIF is not very effective since: 1) existing strategies in this category would fail to account for the joint prior information between the two HRIs, and 2) each of the images can have distinct statistical properties, which makes obtaining adequate priors more difficult. Moreover, although deep image priors are unsupervised~\cite{ulyanov2018deep}, they require careful setup of the network architecture and the number of stochastic gradient iterations to produce reasonable results.
It {must} be noted that these challenges related to the lack of training data and the corresponding difficulty in learning priors of the scene of interest are also encountered more generally in HMIF, i.e., even when inter-image variability is not present.

In this paper, we propose a new image fusion method accounting for inter-image variability between HIs and MIs which addresses the aforementioned challenges. First, to adequately represent the image-specific information as well as the joint prior information between the two HRIs, we propose a mixture distribution that accounts for the leptokurtic nature of the inter-image variations while, at the same time, represents complex image content by implicitly exploiting learning-based image priors. An iteratively reweighted optimization strategy is then proposed, and the regularization by denoising (RED)~\cite{romano2017little} framework is employed to implicitly introduce prior information about the HRIs by means of denoising engines, one for each latent HRI. The denoisers are trained using {a zero-shot strategy~\cite{shocher2018zeroshot}} and adapted during the optimization process, which allows them to account for the content of each individual HRI.
The proposed algorithm is called \emph{Deep hyperspectral and multispectral Image Fusion with Inter-image Variability} (DIFIV). Experiments on data with real inter-image variability demonstrate the superiority of DIFIV compared to other state-of-the-art methods. The contributions of the paper are summarized as follows.
\begin{itemize}
  \item A general imaging model is formulated, 
  where the inter-image variations of the HRIs are modeled by a hyper-Laplacian distribution to account for the joint image content, while the image content specific {to each HRI} is learned by two distinct deep CNNs.
  \item To solve the non-convex, non-smooth HMIF optimization problem,  a variable splitting strategy is combined with an iteratively reweighted scheme to tackle the difficulties introduced by both the hyper-Laplacian and deep priors, which are defined implicitly based on CNN denoisers under the RED framework. 
  
  \item {We use a zero-shot strategy inspired by~\cite{shocher2018zeroshot} to learn the CNN denoisers based only on the observed HI and MI. Moreover, unlike the original use of zero-shot methods for single image restoration, the denoisers are trained iteratively during the optimization process based on the currently estimated HRIs. This allows the learned priors to represent the individual information in each of the HRIs adaptively while incorporating at the same time information from both low resolution images as the method converges.} Furthermore, the architecture of CNNs is made lightweight by considering separable convolutions and a low-rank representation of HIs to yield a small number of network parameters.
  %
\end{itemize}

The paper is organized as follows. In Section~\ref{sec:model_var}, the HI and MI observation processes are presented, as well as a review of recent methods considering inter-image variability. Section~\ref{sec:method} formulates a new model and introduces the proposed method. Experimental results with data containing real inter-image variability are given in Section~\ref{sec:results}. Finally, Section~\ref{sec:con} concludes the paper.



\begin{figure*}[tp] 
	\centering
	\includegraphics[scale=0.5]{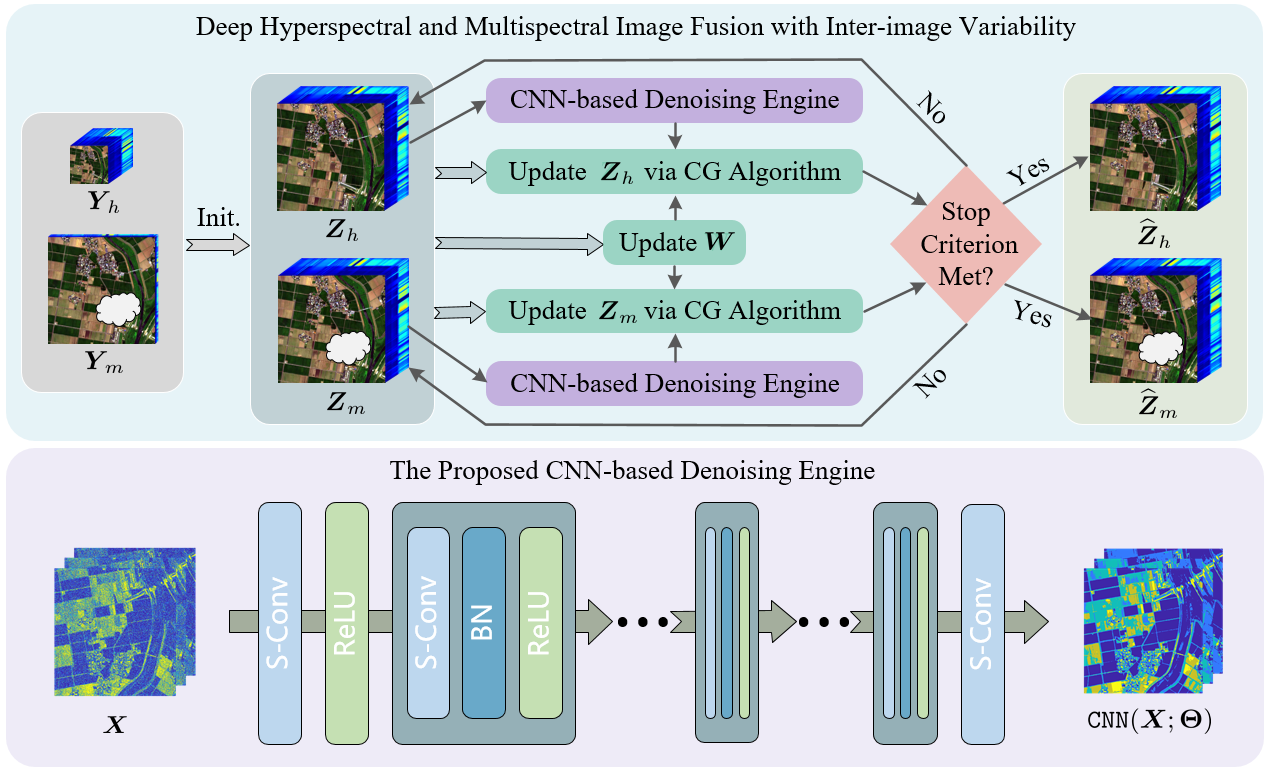}
	\caption{{\textbf{Top panel}: Overall illustration of the proposed Deep hyperspectral and multispectral Image Fusion with Inter-image Variability (DIFIV) method: \cred{the HRIs underlying the HI and MI ($\bZ_h$ and $\bZ_m$) are initialized (Init.), with interpolations of observed images ($\bY_{\!h}$ and $\bY_{\!m}$), and then} used to compute the inter-image variability weighting term $\bW$ and to update the CNN-based denoisers; afterwards, these are used to re-compute the HRIs using a conjugate-gradient based algorithm; this process is repeated iteratively until convergence. \textbf{Bottom panel}: The neural network architecture of our CNN-based denoising engine (S-Conv, BN, and ReLU stand for separable convolution, batch normalization and rectified linear unit layers, respectively).}}
	\label{fig:flowchart}
\end{figure*}

\section{Image Fusion with Inter-image Variability}
\label{sec:model_var}

Let us denote an HI with $L_h$ bands and $N$ pixels by $\bY_{\!h}\in\amsmathbb{R}^{L_h\times N}$, and a MI with $L_m$ bands and $M$ pixels by $\bY_{\!m}\in\amsmathbb{R}^{L_m\times M}$, where $L_m<L_h$ and $N<M$. These images are assumed to be degraded versions of a pair of underlying HRIs $\bZ_h\in\amsmathbb{R}^{L_h\times M}$ and $\bZ_m\in\amsmathbb{R}^{L_h\times M}$ with high spatial and spectral resolutions, which are related according to the following model:
\begin{align}
\begin{split}
    \label{eq:observation_model}
	\bY_{\!h} {}={} & \bZ_h \bF \bD + \bE_h \,,
    \\
    \bY_{\!m} {}={} & \bR \bZ_m + \bE_m \,,
\end{split}
\end{align}
in which matrices $\bF\in\amsmathbb{R}^{M\times M}$ and $\bD\in\amsmathbb{R}^{M\times N}$ represent optical blurring and spatial downsampling occurring at the hyperspectral sensor, respectively; matrix $\bR\in\amsmathbb{R}^{L_m\times L_h}$ contains the spectral response functions (SRF) of the multispectral instrument, and $\bE_h\in\amsmathbb{R}^{L_h\times N}$ and $\bE_m\in\amsmathbb{R}^{L_m\times M}$ denote additive noise.

In this setting, the image fusion problem consists of recovering the HRIs $\bZ_h$ and $\bZ_m$ given the observations $\bY_{\!h}$ and $\bY_{\!m}$. Most of the previous methods consider that $\bY_{\!h}$ and $\bY_{\!m}$ are degraded from the same source, i.e., $\bZ_h=\bZ_m$, which intrinsically assumes that they are acquired under the same conditions, e.g., by sensors on board a single satellite. However, due to the wider availability of satellites equipped with multispectral sensors, it is of great interest to fuse HIs and MIs acquired by different instruments at different time instants~\cite{yokoya2017HS_MS_fusinoRev}.
In that case, by assuming that $\bZ_h=\bZ_m$, most existing methods ignore variabilities between the HI and MI, which can occur due to differences in acquisition conditions caused by, e.g., atmospheric, illumination or seasonal variations~\cite{borsoi2020variabilityReview}, or abrupt changes~\cite{liu2019reviewCD_GRSM}.

Recently, image fusion frameworks addressing inter-image variability have been proposed in~\cite{Borsoi_2018_Fusion,borsoi2020tensorHSRvariability,prevost2022LL1_HSR,borsoi2021tensorHSRasilomar, brezini2021fusionIntraImageVariability,camacho2022HS_MS_fusion_spatial_variability, fu2021fusionInterImageChangesL21norm}. Such methods estimate both HRIs~$\bZ_h$ and~$\bZ_m$ by using different assumptions to model both the images and the inter-image changes.
The first method to address this problem was FuVar~\cite{Borsoi_2018_Fusion}. It considers that the HRIs satisfy the linear mixing model (LMM)~\cite{Keshava:2002p5667}, but with a distinct set of spectral basis vectors for each image:
\begin{align}
    \label{eq:lmm_fuvar}
    \bZ_h=\bM_h\bA \,, \quad \bZ_m = \bM_m\bA\,,
\end{align}
where $\bM_h$ and $\bM_m\in\amsmathbb{R}^{L_h\times R}$ denote the set of spectral basis vectors related to the HI and MI, respectively, and $\bA\in\amsmathbb{R}^{R\times N}$ their corresponding spatial coefficients. {Note that $\bM_h$ and $\bM_m$ are associated with the spectral signatures of the pure materials (i.e., the endmembers) in the HI and MI, respectively.} FuVar considers $\bM_h$ and $\bM_m$ to be related to one another through a set of smooth multiplicative scaling factors $\bPhi\in\amsmathbb{R}^{L_h\times R}$~\cite{imbiriba2018glmm}:
\begin{align}
    \label{eq:model_GLMM_fuvar}
    \bM_m = \bM_h \odot \bPhi \,,
\end{align}
where $\odot$ denotes the Hadamard product. {Thus, this model successfully accounts for changes in the spectral signatures of the endmembers between the HI and the MI, which can occur when the materials are affected by seasonal variations or when the MI is affected by uniform changes caused by, e.g., different illumination conditions.}
%
However, the coefficients $\bA$ shared by both images limit the capability of FuVar to represent inter-image changes in the spatial domain.

This limitation has been addressed by considering spatially and spectrally localized inter-image variations through an additive model in a tensor-based framework~\cite{borsoi2020tensorHSRvariability}. This latter work considers a model of the form:
\begin{align}
    \label{eq:inter_img_var_mdl_cbstar}
    \bZ_h\,, \quad \bZ_m = \bZ_h + \bPsi \,,
\end{align}
where $\bPsi\in\amsmathbb{R}^{L_h\times M}$ denotes a set of additive variability factors. Both the HRI $\bZ_h$ and the variability $\bPsi$ are assumed to admit a Tucker tensor decomposition with low multilinear ranks~\cite{sidiropoulos2017tensor}. This reduces the dimensionality of the problem and allows theoretical identifiability and recovery guarantees to be obtained~\cite{borsoi2020tensorHSRvariability}.

A related work proposes to jointly address the image fusion and hyperspectral unmixing problems in the presence of inter-image spectral variability~\cite{prevost2022LL1_HSR}. This consists of the recovery of both the HRIs and the spectral signatures of the endmembers and their abundances. An LL1 tensor model is considered, which is closely related to the LMM in~\eqref{eq:lmm_fuvar} but involves an additional low-rank assumption on the coefficient maps $\bA$ that allows theoretical identifiability results to be derived. Other works propose to consider intra-image variability by extending the LMM to consider spatial endmember variability, i.e., variability within a single image~\cite{brezini2021fusionIntraImageVariability,camacho2022HS_MS_fusion_spatial_variability}. Another work considers a robust version of the data fidelity term related to the MI in the cost function to reduce the impact of possible changes or outliers in the image fusion process~\cite{fu2021fusionInterImageChangesL21norm}. However, these methods still assume that the HRIs underlying the HI and the MI are equal.


Despite the success of these approaches in addressing the inter-image variability problem, they all rely on handcrafted priors for the HR images $\bZ_h$ and $\bZ_h$, which limits their capability of representing realistic image content.
In this work, we propose an image fusion method that leverages the expressive power of CNNs in order to construct accurate image priors for the HRIs while accounting for inter-image variability, as detailed in the following section.

\section{The proposed method}
\label{sec:method}

The proposed image fusion method is based on three important axes/contributions: 1) an imaging model that incorporates inter-image variability with learned image priors, 2) an optimization scheme that can handle these flexible penalties, 3) a lightweight unsupervised (zero-shot) scheme to iteratively learn deep priors of the latent HRIs during the reconstruction process.
The proposed image fusion method is presented through four steps. First, we present the imaging model in Subsection~\ref{ssec:our_img_mdl} and formulate the optimization problem. In Subsection~\ref{ssec:iter_reweight} we describe an iteratively reweighted scheme to optimize the cost function. The optimization steps, as well as the integration of deep priors, are described in Subsection~\ref{ssec:optimization}. We then address the design of CNN architecture and its image-adapted training strategy in Subsection~\ref{ssec:learn_deep_prior}. {An overall illustration of the proposed DIFIV method is shown in Figure~\ref{fig:flowchart}.}

\subsection{The imaging model}
\label{ssec:our_img_mdl}
Using a probabilistic framework, the HMIF problem can be formulated as the recovery of the mean or mode of the posterior probability distribution function (PDF) $p(\bZ_h,\bZ_m|\bY_{\!h},\bY_{\!m})$ of both HRIs given the LR observations. Using Bayes theorem, this PDF can be written as:
\begin{align} \label{eq:posterior_distro}
    p(\bZ_h,\bZ_m|\bY_{\!h},\bY_{\!m}) \,\,\propto\,\,
    & p(\bY_{\!h}|\bZ_{h})p(\bY_{\!m}|\bZ_{m})
    \nonumber \\
    & p(\bZ_m,\bZ_h) \,,
\end{align}
where we assumed the HI and MI to be conditionally independent given their high-resolution counterparts.

The likelihoods of the observed images $\bY_{\!h}$ and $\bY_{\!m}$ can be written according to their data generation process in~\eqref{eq:observation_model}. More precisely, assuming the elements of $\bE_h$ and $\bE_m$ to be i.i.d. Gaussian random variables with variance $\sigma_h^2$ and $\sigma_m^2$, respectively, the conditional distributions of $\bY_m$ and $\bY_n$ in~\eqref{eq:posterior_distro} are given by:
\begin{align}
    p(\bY_{\!h}|\bZ_{h}) ={} & \calM\calN(\bZ_h\bF\bD, \sigma_h^2\bI_{L_h}, \bI_{N}) \,,
    \\
    p(\bY_{\!m}|\bZ_{m}) ={} & \calM\calN(\bR\bZ_m, \sigma_m^2\bI_{L_m}, \bI_{M}) \,,
\end{align}
where $\calM\calN(\bUpsilon, \bSigma_r, \bSigma_c)$ denotes the matrix normal distribution with mean matrix $\bUpsilon$ and row and column covariance matrices $\bSigma_r$ and $\bSigma_c$, respectively~\cite{wei2015hs_ms_fusionSylvesterEq}.

The challenging question concerns how to meaningfully define the joint prior $p(\bZ_m,\bZ_h)$ for both HRIs. This question is not trivial when the images differ due to acquisition conditions or seasonal variations. A simplistic possibility is to consider the images to be independent and to use priors used for super-resolution without variability, such as low-rank matrix and tensor models~\cite{yokoya2012coupledNMF,kanatsoulis2018hyperspectralSRR_coupledCPD,prevost2019coupledTucker_hyperspectralSRR}, piecewise-smoothness~\cite{simoes2015HySure} or learned deep priors~\cite{wang2021hyperspectral, dian2020regularizing, wang2022hyperspectral}.
However, the images $\bZ_m$ and $\bZ_h$ are observations of the same scene, and thus are strongly dependent. Considering this, we can state the following desirable properties for $p(\bZ_m,\bZ_h)$:
\begin{itemize}
    \item Apart from possible smooth inter-image variations (such as, e.g., illumination or atmospheric changes, which tend to impact the images uniformly~\cite{borsoi2020variabilityReview}), changes between $\bZ_m$ and $\bZ_h$ are generally small and sparse; high magnitude changes are concentrated in a relatively small number of pixels and bands~\cite{liu2019reviewCD_GRSM}.
    
    \item {The prior should promote} images $\bZ_m$ and $\bZ_h$ which are statistically similar to real hyperspectral images (e.g., they can be well represented by learned priors).
    
\end{itemize}


To achieve the above desiderata, we consider a mixture distribution, given by:
\begin{align}
    \log p(\bZ_m,\bZ_h) {}\,\,\propto\,\,{} & - \frac{\lambda_p}{2}
    \sum_{\ell,n}  \big|\delta_h^{(\ell,n)}-\delta_m^{(\ell,n)}\big|^p
    \nonumber\\
    & - \lambda_m \phi_m(\bZ_m) - \lambda_h \phi_h(\bZ_h) \,,
    \label{eq:prior_Zh_Zm}
\end{align}
for $0<p\leq 1$,
where $\delta_h^{(\ell,n)}$ and $\delta_m^{(\ell,n)}$ denote the $(\ell,n)$-th locations of a high-pass spatio-spectral filtered version of $\bZ_h$ and $\bZ_m$, which are denoted by $\bDelta_h$ and $\bDelta_m$, respectively. We assume this filtering to be computed through an operator $\calG$ satisfying $\bDelta_h=\calG(\bZ_h)$, $\bDelta_m=\calG(\bZ_m)$, and in vector form as $\vect(\bDelta_h)=\bG\vect(\bZ_h)$ and $\vect(\bDelta_m)=\bG\vect(\bZ_m)$ where $\bG$ is the matrix form of $\calG$. One natural example for $\calG$ is the spatio-spectral gradient operator, e.g. Laplacian filter. Parameters $\lambda_p$, $\lambda_m$ and $\lambda_h$ are regularization parameters.

The first term in~\eqref{eq:prior_Zh_Zm} corresponds to an i.i.d. hyper-Laplacian distribution for the difference between the filtered HRIs~\cite{krishnan2009deconvolutionHyperLaplacian}, which has also been previously used to represent the gradient of the HRI in image fusion~\cite{peng2021HSR_hyperLaplacian}. This distribution is effective for modeling leptokurtic (i.e., heavy-tailed) distributions such as images~\cite{krishnan2009deconvolutionHyperLaplacian}. This can represent an important characteristic of the inter-image changes since these can be restricted to a comparatively small number of pixels and are concentrated at low-frequency spatial content~\cite{borsoi2018newSRRinnovations}.
The functions $\phi_h(\cdot)$ and  $\phi_m(\cdot)$ encode prior knowledge about each HRI, and will be learned implicitly by using deep CNNs.

{Note that the prior in~\eqref{eq:prior_Zh_Zm} also corresponds to a model for the inter-image variability, which can be written as:
\begin{align}
    \label{eq:inter_img_prior_proposed}
    \bZ_m = \bZ_h + \bPsi_{\Delta} \,.
\end{align}
What is distinctive in~\eqref{eq:inter_img_prior_proposed} when compared to the model in~\eqref{eq:inter_img_var_mdl_cbstar} is how prior information is chosen. The prior for the inter-image variability term $\bPsi_{\Delta}$ cannot be written in an analytical form; instead, its properties follow from the interactions of the different terms in~\eqref{eq:prior_Zh_Zm}. The first term encourages the inter-image variability $\bPsi_{\Delta}$ to have small and sparse gradients. The last two terms employ CNNs that can incorporate realistic prior information about each of the HRIs, and only constrain $\bPsi_{\Delta}$ indirectly through its effect on $\bZ_m$ and $\bZ_h$.}

Given this model, the image fusion problem then consists of finding the HRIs $\bZ_h$ and $\bZ_m$ which maximize the logarithm of the posterior distribution $p(\bZ_h,\bZ_m|\bY_{\!h},\bY_{\!m})$ defined in~\eqref{eq:posterior_distro}. 
This corresponds to the following optimization problem:
\begin{align}
     \mathop{\min}_{\bZ_h, \bZ_m} &\,\, \frac{1}{2}\|\bY_{\!h} - \bZ_h\bF\bD\|_F^2 + \frac{1}{2} \|\bY_{\!m} - \bR \bZ_m\|_F^2
    \nonumber\\&
    + \lambda_h \phi_h(\bZ_h) + \lambda_m \phi_m(\bZ_m) \nonumber
    \\&
    + \frac{\lambda_p}{2} \|\calG(\bZ_h) - \calG(\bZ_m)\|_{p,p}^p \,,
    \label{eq:cost_function1}
\end{align}
where $\|\cdot\|_{p,p}$ is the entrywise $L_p$ matrix norm, satisfying $\|\calG(\bZ_h) - \calG(\bZ_m)\|_p^p=\sum_{\ell,n} \big|\delta_h^{(\ell,n)}-\delta_m^{(\ell,n)}\big|^p$. The spatial and spectral priors of $\bZ_m$ and $\bZ_h$ are encoded in $\phi_h(\bZ_h)$ and $\phi_m(\bZ_m)$, respectively.

\subsection{An iteratively reweighted update scheme}
\label{ssec:iter_reweight}

Optimizing the cost function in~\eqref{eq:cost_function1} is challenging. Apart from the image priors $\phi_h(\cdot)$ and $\phi_m(\cdot)$ that will be defined in the sequel, the inter-image prior term (i.e., the last term in~\eqref{eq:cost_function1}) is, in general, a non-convex and non-smooth function of both $\bZ_h$ and $\bZ_m$, which is not straightforward to optimize. To address this problem, we consider an iteratively reweighted optimization strategy~\cite{lu2014iterativeReweightedOptim, ammanouil2014blind}.
First, note that the last term in~\eqref{eq:cost_function1} can be written as:
\begin{align}
    \sum_{\ell,n}  \big|\delta_h^{(\ell,n)}-\delta_m^{(\ell,n)}\big|^p
    = \sum_{\ell,n} w_{\ell,n} \big|\delta_h^{(\ell,n)}-\delta_m^{(\ell,n)}\big|^2  \,,
    \label{eq:joint}
\end{align}
where the weights $w_{\ell,n}$ are given by
\begin{align}
    w_{\ell,n}=|\delta_h^{(\ell,n)}-\delta_m^{(\ell,n)}|^{p-2}  \,.
    \label{eq:w}
\end{align}
Since $w_{\ell,n}\geq0$,~\eqref{eq:joint} can be expressed as:
\begin{align}
    \sum_{\ell,n} w_{\ell,n} \big|\delta_h^{(\ell,n)}-\delta_m^{(\ell,n)}\big|^2 = \|\bW\odot(\bDelta_h-\bDelta_m)\|_F^2 \,,
    \label{eq:inter_im_w}
\end{align}
where $\bW$ is a matrix whose $(\ell,n)$-th entry is given by $\sqrt{w_{\ell,n}}$, and $\odot$ denotes the Hadamard product.

When  matrix $\bW$ is fixed,~\eqref{eq:inter_im_w} becomes a quadratic function of the HRIs, which can be effectively optimized. The nonlinear dependency of $\bW$ on $\bZ_h$ and $\bZ_m$ will be resolved by using an iterative strategy: first the cost function is optimized considering $\bW$ fixed to obtain $\bZ_h$ and $\bZ_m$, and afterwards $\bW$ is updated according to an approximate version of~\eqref{eq:w} by using the values of $\bZ_h$ and $\bZ_m$ computed from previous iteration~\cite{lu2014iterativeReweightedOptim}. This leads to the following iterative procedure, which is repeated until convergence:\\[0.1cm]
1) For a fixed $\bW$, compute $\bZ_h$ and $\bZ_m$ by solving the following optimization problem:
\begin{align}
     \mathop{\min}_{\bZ_h, \bZ_m, \bDelta_h, \bDelta_m} &\,\, \frac{1}{2}\|\bY_{\!h} - \bZ_h\bF\bD\|_F^2 + \frac{1}{2} \|\bY_{\!m} - \bR \bZ_m\|_F^2
    \nonumber\\&
     + \lambda_h \phi_h(\bZ_h) + \lambda_m \phi_m(\bZ_m) 
    \nonumber \\&
    + \frac{\lambda_p}{2}\|\bW\odot(\bDelta_h-\bDelta_m)\|_F^2
    \label{eq:prob_smoothed}
    \\
    \text{s.t.} & \,\,\, \bDelta_h=\calG(\bZ_h),\, \bDelta_m=\calG(\bZ_m) \,. 
    \nonumber
\end{align}
2) Update the entries of $\bW$ according to
\begin{align}
    w_{\ell,n}=\big(|\delta_h^{(\ell,n)}-\delta_m^{(\ell,n)}|+\epsilon\big)^{p-2}  \,,
    \label{eq:w_epsilon}
\end{align}
where $\epsilon>0$ is a small constant included in~\eqref{eq:w} to ensure the numerical stability of the algorithm.\\[0.1cm]
3) Return to step 1) and repeat until convergence.




This strategy is efficient to solve sparsity-regularized optimization problems~\cite{daubechies2010iterativelyReweightedSparseLS}. \cmag{Moreover, iteratively reweighted optimization schemes have been shown to converge to a local stationary point under relatively mild conditions~\cite{lu2014iterativeReweightedOptim}.}

In the following subsection, we shall focus on the minimization problem~\eqref{eq:prob_smoothed}.

\subsection{The optimization problem}
\label{ssec:optimization}
Handcrafting powerful regularizers $\phi_h(\bZ_h)$ and $\phi_m(\bZ_m)$ along with solving the associated optimization problems efficiently is not a trivial task. In this subsection, we propose to learn the image prior directly from the observed data and incorporate it into the model-based optimization~\eqref{eq:prob_smoothed} to avoid designing regularizers analytically.

%
First, by introducing two auxiliary variables, $\bV_{\!h}=\bZ_h$ and $\bV_{\!m}=\bZ_m$, problem~\eqref{eq:prob_smoothed} can be rewritten equivalently as:
\begin{align}
    \mathop{\min}_{\Omega} \,\,\, & \frac{1}{2}\|\bY_{\!h} - \bZ_h\bF\bD\|_F^2 
    \nonumber 
    +  \frac{1}{2} \|\bY_{\!m} - \bR \bZ_m\|_F^2 
    \\ & + \frac{\lambda_p}{2}\|\bW\odot(\bDelta_h-\bDelta_m)\|_F^2 
    + \lambda_m\phi_m(\bV_{\!m}) 
    \label{eq:prob_smoothed_2}\\&
    + \lambda_h\phi_h(\bV_{\!h}) 
    \nonumber \\
    \text{s.t.} & \quad \bV_{\!h}=\bZ_h \,,\, \bV_{\!m}=\bZ_m \,, 
    \nonumber \\ &
    \quad\, \bDelta_h=\calG(\bZ_h)\,,\, \bDelta_m=\calG(\bZ_m) \,,
    \nonumber
\end{align}
where $\Omega=\{\bZ_h,\bZ_m,\bV_{\!h},\bV_{\!m},\bDelta_h,\bDelta_m\}$. By using the half-quadratic splitting (HQS) approach~\cite{geman1995nonlinear}, we can decouple the data fidelity and regularization terms in~\eqref{eq:prob_smoothed_2} and write this cost function as:
\begin{align}
   & \mathcal{L}_{\rho}(\bZ_h,\bZ_m,\bV_{\!h},\bV_{\!m}) = 
    \frac{1}{2}\|\bY_{\!h} - \bZ_h\bF\bD\|_F^2 \nonumber \\
   & +  \frac{1}{2} \|\bY_{\!m} - \bR \bZ_m\|_F^2 + \frac{\lambda_p}{2}\|\bW\odot(\calG(\bZ_h) - \calG(\bZ_m))\|_F^2 
   \nonumber\\&
   + \frac{\rho}{2}\|\bZ_{m} - \bV_{\!m}\|_F^2 +\frac{\rho}{2}\|\bZ_{h} - \bV_{\!h}\|_F^2
    \\
    & + \lambda_m\phi_m(\bV_{\!m})
    + \lambda_h\phi_h(\bV_{\!h}) \,, \nonumber
\end{align}
with $\rho\in\amsmathbb{R}_+$ the penalty parameter. In the following, we consider a block coordinate descent (BCD) strategy and minimize $\mathcal{L}_{\rho}$ with respect to each variable, one at a time.

\textbf{Optimization w.r.t. $\bZ_h$:} This optimization problem can be written as:
\begin{align}
\label{sub_z_h}
    \min_{\bZ_h} \,\,\, & \frac{1}{2} \|\bY_{\!h} - \bZ_h\bF\bD\|_F^2 + \frac{\lambda_p}{2}\|\bW\odot(\calG(\bZ_h) - \calG(\bZ_m))\|_F^2 
    \nonumber\\&
    + \frac{\rho}{2}\|\bZ_h-\bV_{\!h}\|_F^2 \,.
\end{align}
By taking the derivative of the cost function in~\eqref{sub_z_h}, setting it equal to zero and using the vectorization property of matrix products, we obtain:
\begin{align}
    & - \big[(\bF\bD)^\top\otimes\bI_L\big]^\top\Big(\vect(\bY_{\!h}) - \big[(\bF\bD)^\top\otimes\bI_L\big]\vect(\bZ_h)\Big) 
    \nonumber \\ &
    + \lambda_p\bG^\top\diag(\vect(\bW))^2\bG \big(\vect(\bZ_h-\bZ_m)\big)
    \nonumber \\ &
    + \rho\vect(\bZ_h-\bV_{\!h}) = \cb{0} \,.
\end{align}
Using the properties of the Kronecker product, this equation can be written as:
\begin{align}
    & \Big(\big[(\bF\bD)(\bF\bD)^\top\otimes\bI_L\big] + \lambda_p\bG^\top\diag(\vect(\bW))^2\bG
    \nonumber\\ &
    + \rho\bI \Big)\vect(\bZ_h)
    =  \big[(\bF\bD)^\top\otimes\bI_L\big]^\top\vect(\bY_{\!h}) 
    \nonumber\\ &
    + \lambda_p\bG^\top\diag(\vect(\bW))^2\bG \vect(\bZ_m)
    + \rho\vect(\bV_{\!h}) \,.
    \label{eq:z_h}
\end{align}
which is a linear system of equations. However, solving this system directly is prohibitive due to its large dimension. Since the matrix on the left-hand side is symmetric positive-definite, we propose to solve this problem using the conjugate gradient (CG) algorithm, which requires only matrix-vector products that can be implemented implicitly and more efficiently.

\textbf{Optimization w.r.t. $\bZ_m$:} This optimization problem can be written as:
\begin{align}
    \min_{\bZ_m} \,\,\, & \frac{1}{2} \|\bY_{\!m} - \bR \bZ_m\|_F^2 + \frac{\lambda_p}{2}\|\bW\odot(\calG(\bZ_h) - \calG(\bZ_m))\|_F^2
    \nonumber\\&
    + \frac{\rho}{2}\|\bV_{\!m}-\bZ_m\|_F^2 \,.
\end{align}
Following the same steps as for problem~\eqref{sub_z_h}, we obtain:
\begin{align}
    & \Big(\big[\bI_N\otimes\bR^\top\bR\big] + \lambda_p\bG^\top\diag(\vect(\bW))^2\bG +\rho\bI\Big)\vect(\bZ_m)
    \nonumber\\ &
    = \big[\bI_N\otimes\bR\big]^\top\vect(\bY_{\!m}) + \lambda_p\bG^\top\diag(\vect(\bW))^2\bG \vect(\bZ_h)
    \nonumber\\ &
    + \rho\vect(\bV_{\!m}) \,.
    \label{eq:z_m}
\end{align}
Considering that the matrix on the left-hand side is symmetric positive-definite, the CG algorithm is used to solve this problem.


\textbf{Optimization w.r.t. $\bV_{\!h}$:} 
This optimization problem can be written as:
\begin{align}
    \min_{\bV_{\!h}} \,\,\, \frac{\rho}{2}\|\bV_{\!h}-\bZ_h\|_F^2 + \lambda_h\phi_h(\bV_{\!h}) \,.
    \label{eq:v_h_opt}
\end{align}
As discussed above, designing accurate handcrafted regularizers for $\phi_h(\bV_{\!h})$ may be complicated. To address this issue efficiently, we propose to use a strategy that leverages a CNN denoiser. Popular strategies are the Plug-and-Play (PnP) framework~\cite{venkatakrishnan2013plug} and the Regularization by Denoising (RED) scheme~\cite{romano2017little}. 
{In this work, we consider the RED strategy since it is associated with an explicit optimization objective and because it was experimentally shown in~\cite{romano2017little} to have more stable convergence and robustness in relation to the selection of hyperparameters when compared to PnP methods.}
Consider denoising an HI $\bV$, we define the CNN denoiser as $\calD(\bV)$.
RED framework defines $\phi_h(\cdot)$ as the inner product between an image and its denoising residual:
\begin{align}
    \phi_h(\bV) = \frac{1}{2}\langle \bV, \bV-\calD(\bV)\rangle \,,
    \label{eq:RED_reg_def}
\end{align}
where $\langle\cdot, \cdot\rangle$ denotes the inner product. This can be interpreted as an image-adaptive Laplacian regularizer.
Using~\eqref{eq:RED_reg_def}, the optimization problem~\eqref{eq:v_h_opt} becomes 
\begin{align}
    \min_{\bV_{\!h}} \,\,\, \frac{\rho}{2}\|\bV_{\!h}-\bZ_h\|_F^2 + \frac{\lambda_h}{2}\langle\bV_{\!h}, \bV_{\!h} - \calD(\bV_{\!h})\rangle \,.
\end{align}
Taking the derivative of the cost function and setting it to zero, we obtain:
\begin{align}
    \rho(\bV_{\!h}-\bZ_h) + \lambda_h(\bV_{\!h} - \calD(\bV_{\!h})) = \cb{0} \,.
\end{align}
To solve this equation, a fixed-point iterative update is used, leading to the following recursive update equation:
\begin{align}
    \bV_{\!h}^{(i+1)} = \frac{1}{\rho+\lambda_h}\big(\rho\bZ_h+ \lambda_h\calD(\bV_{\!h}^{(i)})\big) \,.
    \label{eq:v_h}
\end{align}
where $\bV_{\!h}^{(i)}$ denotes the solution $\bV_{\!h}$ at the $i$-th iteration.




\textbf{Optimization w.r.t. $\bV_{\!m}$:} 
Following the same strategy as above, we obtain:
\begin{align}
    \bV_{\!m}^{(i+1)} = \frac{1}{\rho+\lambda_m}\big(\rho\bZ_m+ \lambda_m\calD(\bV_{\!m}^{(i)})\big) \,.
    \label{eq:v_m}
\end{align}
Note that we only use {a single step for the fixed point} iteration in~\eqref{eq:v_h} and~\eqref{eq:v_m} for computational efficiency.

\subsection{Learning deep priors via image-specific CNNs}
\label{ssec:learn_deep_prior}
Generally, function $\calD(\, \cdot\,)$ can be any off-the-shelf denoiser. This offers the opportunity of incorporating a fast CNN denoising engine with powerful prior learning ability into physical model-based iterative optimization procedure~\cite{chen2022integration}. However, there are three main challenges in using CNN denoisers to learn priors for hyperspectral images in RED or PnP frameworks~\cite{dian2020regularizing, wang2020learning}: First, there is a {limited amount of data available for training}; second, there is an even greater {scarcity of labeled training data}; third, the {noise level of the HRI to be denoised in~\eqref{eq:v_h}-\eqref{eq:v_m} changes} over the BCD iterations as the method converges. To overcome each of these challenges, we propose a lightweight, unsupervised and image-specific CNN denoiser, which is detailed in the following.


\textbf{Lightweight network architecture:} To overcome the limited number of available data to train efficient CNN denoisers, a lightweight architecture with fewer parameters needs to be considered in the network design. In this work, two strategies have been considered to lighten network architecture, namely: 1) dimensionality reduction of the input image, which reduces the number of CNN filters, and 2) separable convolutions~\cite{howard2017mobilenets}, which reduces the filter volume (i.e., the number of parameters of each filter).

We considered the DnCNN~\cite{zhang2017beyond} as a backbone in network design.
For color (i.e., RGB) images, each layer of DnCNN contains $64$ filters. Directly using this network architecture to denoise an HI $\bV$ with $L_h$ channels would approximately lead to the use of $64 \times L_h / 3$ filters in each layer, leading to a very large number of parameters.
This increase in the number of network parameters makes it hard to train since the amount of training data is usually very limited. Considering that the spectral channels of $\bV$ are highly correlated and contain highly redundant information, we can assume that there exists a subspace of dimension much lower than $L_h$ which captures all the information of $\bV$. This allows us to write $\bV$ using a low-rank representation as:
\begin{align}
    \bV =  \bQ\bX \,,
    \label{eq:svd}
\end{align}
where $\bQ\in\amsmathbb{R}^{L_h\times l_h}\ (l_h \ll L_h, \bQ^\top\bQ = \bI_{l_h})$ and $\bX\in\amsmathbb{R}^{l_h\times M}$ are the subspace matrix and the representation coefficients, respectively. Small values of $l_h$ correspond to data description in a low-dimensional space. 
Employing such dimensionality reduction in the CNN denoising engine has a core benefit. It decreases the number of filters by a ratio of $l_h / L_h$ in each layer by removing the burden of learning information that is redundant across spectral channels. 

To reduce filter volume, we use separable convolutions to further lighten the backbone architecture as in~\cite{imamura2019zero}. In particular, the core idea of separable convolution is decomposing a convolution filter with $3\times 3 \times \texttt{Depth}$ parameters into a depth-wise filter with $3\times 3 \times 1$ parameters and a point-wise filter with $1\times 1 \times \texttt{Depth}$ parameters, where $\texttt{Depth}$ is the input depth of this CNN layer. This reduces the number of parameters by a rate of $1 / \texttt{Depth} + 1/ (3\times3)$. 
Thus, the lightweight DnCNN contains three kinds of operators: $3\times 3$ separable convolution layers (S-Conv), rectified linear units (ReLU) and batch normalization (BN).  ReLU is the activation function while BN is used to accelerate the training speed. In the network architecture, the first layer is ``S-Conv + ReLU'', the hidden layer is ``S-Conv + BN + ReLU'' and the last layer is ``S-Conv''. {This network architecture is illustrated in the bottom panel of Figure~\ref{fig:flowchart}.}  Furthermore, we adopt the residual learning strategy in~\cite{zhang2017beyond} to predict the residual image before achieving the estimated clean image.

With these two strategies, the number of network parameters can be significantly reduced with a ratio of $(l_h / L_h)\times(1 / \texttt{Depth} + 1/ (3\times3))$, which is key to allowing the denoising engine to learn a powerful prior from a small training set.

\textbf{Zero-shot training strategy:} In many real-world scenarios, training data with paired noisy and clean images related to the scene of interest are not available. Moreover, using synthetic training data or images from different sites may lead to the so-called domain shift, where the model does not perform well due to differences between the statistical distribution of training and test data~\cite{dian2020regularizing, wang2020learning}. 
{Therefore, it is desirable to consider a training strategy that is \emph{zero-shot}, that is, which is unsupervised and uses only the information of the observed noisy HI and MI pair} itself for training.

Thus, we propose to leverage the information inside a single image to train the CNN denoiser. Natural images have significant information redundancy across different spatial positions and scales, which has been successfully exploited in single image restoration algorithms~\cite{glasner2009multiscaleSRR}. 
Consider the CNN-based denoiser $\texttt{CNN}(\, \cdot\, ; {\Theta})$ with network parameters $\Theta$, and an observed noisy image $\bX$ generated following the degradation model $\bX=\bZ+\bE$, where $\bE$ is i.i.d. Gaussian noise with a standard deviation $\sigma$. $\texttt{CNN}(\, \cdot\, ; {\Theta})$. To learn the CNN denoiser $\texttt{CNN}(\, \cdot\, ; {\Theta})$, we make the important assumption that the set of parameters $\Theta$ which allow it to recover $\bZ$ from $\bX$, are the same as those which allow $\texttt{CNN}(\, \cdot\, ; {\Theta})$ to recover $\bX$ from $\bX+\bE$. This assumption has been used to learn image-adapted CNNs for super-resolution in~\cite{shocher2018zeroshot}. It allow us to train the denoising engine $\texttt{CNN}(\, \cdot\, ; {\Theta})$ using the image pair $(\bX + \bE, \bX)$ by minimizing the following $\ell_1$-norm loss function:
\begin{equation}\label{eq:loss}
    \ell(\Theta) = \|\texttt{CNN}(\bX + \bE;\Theta)-\bX\|_{1} \,.
\end{equation}
Note that the noisy-clean image pair $(\bX + \bE, \bX)$ is generated by adding Gaussian noise with standard deviation $\sigma$ to the observation $\bX$. We adopted the method described in~\cite{donoho1994ideal} to estimate $\sigma$ in each channel of $\bX$. 

The procedure for learning the proposed CNN-based denoising engine is summarized in Algorithm~\ref{alg_cnn}. 
\cred{Note that the training procedure considers the entire image, $\bX$. However, for large images, other learning objectives that decompose the image into different patches or across multiple scales can provide ways to parallelize the training procedure, which might reduce the execution times.}

\textbf{Image-specific prior learning:} Since there exist some inter-image variations between $\bZ_h$ and $\bZ_m$, we considered to train two independent denoising engines $\texttt{CNN}(\, \cdot\, ; {\Theta_h})$ and $\texttt{CNN}(\, \cdot\, ; {\Theta_m})$ to denoise $\bV_{\!h}$ and $\bV_{\!m}$, respectively. This leads to different denoising engines, which can be expressed by substituting $\calD$ by $\calD_h$ in~\eqref{eq:v_h}, and by $\calD_m$ in~\eqref{eq:v_m}.

In general, the equivalent noise levels of $\bV_{\!h}$ and $\bV_{\!m}$ decrease over the BCD iterations since the reconstructed images get closer to the ground truth. Thus, $\texttt{CNN}(\, \cdot\, ; {\Theta_h})$ and $\texttt{CNN}(\, \cdot\, ; {\Theta_m})$ should have the ability to tackle multiple noise levels. To address this issue, we propose a strategy that adaptively updates network parameters $\Theta_h$ and $\Theta_m$ to learn an image-specific prior at each BCD iteration. This is performed by re-training $\texttt{CNN}(\, \cdot\, ; {\Theta_h})$ and $\texttt{CNN}(\, \cdot\, ; {\Theta_m})$ to denoise the estimates of the HRIs at the current BCD iteration. To make the algorithm faster, we consider training $\texttt{CNN}(\, \cdot\, ; {\Theta_h})$ and $\texttt{CNN}(\, \cdot\, ; {\Theta_m})$ in the first BCD iteration and then fine-tune them in all the remaining iterations.

Overall, after overcoming the discussed challenges with the above strategies, the denoising engine in Algorithm~\ref{alg_cnn} is incorporated into the model-based optimization procedure described in Subsection~\ref{ssec:optimization}. The overall DIFIV strategy is described in Algorithm~\ref{alg}.

\renewcommand{\algorithmicrequire}{ \textbf{Input:}} 
\renewcommand{\algorithmicensure}{ \textbf{Output:}} 
\begin{algorithm}[!t]
	\caption{The Proposed CNN-based denoising engine.}
	\label{alg_cnn}
	\begin{algorithmic}
		\REQUIRE Noisy image $\bV$ and subspace dimension $l_h$.\\\vspace{1mm}
		\ENSURE Denoised image $\calD(\bV)$.\\
		\STATE Find $\bQ$ and $\bX$ in~\eqref{eq:svd} using the (truncated) SVD of $\bV$.
		\STATE Optimize $\Theta$ by minimizing~\eqref{eq:loss} with back-propagation.
		\vspace{1mm}
		\STATE Denoise $\bX$ with $\Theta$ as $\texttt{CNN}(\bX; {\Theta})$.
		\vspace{1mm}
		\STATE Transform $\texttt{CNN}(\bX; {\Theta})$ to $\calD(\bV) = \bQ\,\texttt{CNN}(\bX; {\Theta})$.
	\end{algorithmic}
\end{algorithm}

\renewcommand{\algorithmicrequire}{ \textbf{Input:}} 
\renewcommand{\algorithmicensure}{ \textbf{Output:}} 
\begin{algorithm}[!t]
	\caption{Deep Hyperspectral and Multispectral Image Fusion with Inter-image Variability (DIFIV).}
	\label{alg}
	\begin{algorithmic}
		\REQUIRE $\bY_{\!h}, \bY_{\!m}, \bF, \bD, \bR$, paramters $p, \lambda_p, \lambda_h, \lambda_m, \rho$.\\\vspace{1mm}
		\ENSURE The estimated high-resolution images ${\bZhat}_{h}, {\bZhat}_{m}$.\\
		\STATE Interpolate $\bY_{\!h}$ and $\bY_{\!m}$ as $\widetilde{\bY}_{\!h}$ and $\widetilde{\bY}_{\!m}$, respectively.
		\vspace{0.5mm}
		\STATE Initialize $\bZ_h = \bV_{\!h}=\widetilde{\bY}_{\!h}$ and $\bZ_m = \bV_{\!m}=\widetilde{\bY}_{\!m}$.
		\STATE Initialize $\bW$ using~\eqref{eq:w}.
		\WHILE{stopping criteria are not met}
		\STATE Calculate $\bZ_h$ by solving~\eqref{eq:z_h} via CG algorithm.
		\STATE Calculate $\bZ_m$ by solving~\eqref{eq:z_m} via CG algorithm.
		\STATE Update $\bW$ using~\eqref{eq:w}.
		\STATE Learn deep priors via denoising $\bV_{\!h}$ with Algorithm~\ref{alg_cnn}.
		\STATE Update $\bV_{\!h}$ via~\eqref{eq:v_h}.
		\STATE Learn deep priors via denoising $\bV_{\!m}$ with Algorithm~\ref{alg_cnn}.
		\STATE Update $\bV_{\!m}$ via~\eqref{eq:v_m}.
		\ENDWHILE
	\end{algorithmic}
\end{algorithm}

\section{Experiments}
\label{sec:results}

In this section, the effectiveness of the proposed DIFIV method is illustrated through numerical experiments considering two categories of real data, i.e., observed images with moderate and significant inter-image variability. The results provided by the DIFIV are compared with other state-of-the-art hyperspectral and multispectral image fusion methods from both quantitative and qualitative perspectives. The code is made available at \url{https://github.com/xiuheng-wang/DIFIV_release}.

\subsection{Experimental setup} \label{sec:sim_setup}
\begin{figure}[tb]  \footnotesize
	\centering
    \begin{minipage}{0.9\linewidth}
        \centering
        \includegraphics[scale = 0.7]{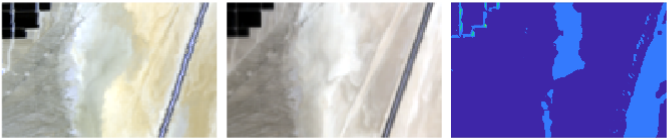}
        Ivanpah Playa
        \vspace{2mm}
    \end{minipage}
    \,\,
    \begin{minipage}{0.9\linewidth}
        \centering
        \includegraphics[scale = 0.7]{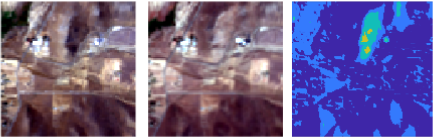}
        \\Lake Isabella
        \vspace{2mm}
    \end{minipage}
    \,\,
    \begin{minipage}{0.9\linewidth}
        \centering
        \includegraphics[scale = 0.7]{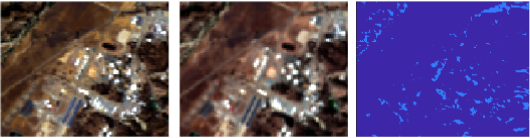}
        \\Lockwood
        \vspace{2mm}
    \end{minipage}
            \,\,
    \begin{minipage}{0.9\linewidth}
        \centering
        \includegraphics[scale = 0.7]{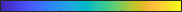}
        \\0 \qquad \ \ 0.2 \qquad \ \ 0.4 \qquad \ \ 0.6
    \end{minipage}
    \caption{Visible representation {of the} hyperspectral {(left panels)} and multispectral images {(middle panels)} with moderate variability used in the experiments {and their inter-image changes maps (right panels).}}
    \label{fig:HSIs_MSIs_a}
\end{figure}

\begin{figure}[tb] \footnotesize
	\centering
    \begin{minipage}{0.9\linewidth}
        \centering
        \includegraphics[scale = 0.7]{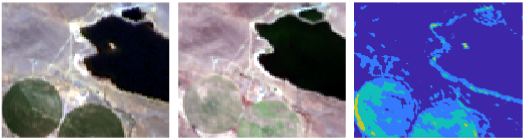}
        \\Lake Tahoe A
        \vspace{2mm}
    \end{minipage}
    \,\,
    \begin{minipage}{0.9\linewidth}
        \centering
        \includegraphics[scale = 0.7]{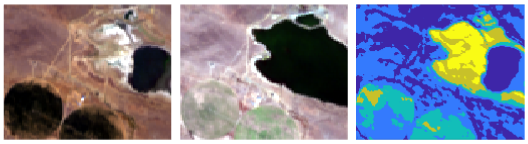}
        \\Lake Tahoe B
    \end{minipage}
        \,\,
    \begin{minipage}{0.9\linewidth}
        \centering
        \vspace{2mm}
        \includegraphics[scale = 0.85]{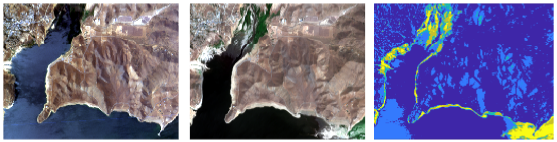}
        \\Kern River
         \vspace{2mm}
    \end{minipage}
            \,\,
        \begin{minipage}{0.9\linewidth}
        \centering
        \includegraphics[scale = 0.7]{figures/colorbar.png}
        \\0 \qquad \ \ 0.2 \qquad \ \ 0.4 \qquad \ \ 0.6
    \end{minipage}
    \caption{Visible representation {of the} hyperspectral {(left panels)} and multispectral images {(middle panels)} with significant variability used in the experiments {and their inter-image changes maps (right panels).}}
    \label{fig:HSIs_MSIs_b}
\end{figure}

We compared our method to nine other techniques, namely: the matrix factorization-based methods HySure~\cite{simoes2015HySure} and CNMF~\cite{yokoya2012coupledNMF}, tensor-based image fusion methods  STEREO~\cite{kanatsoulis2018hyperspectralSRR_coupledCPD} and SCOTT~\cite{prevost2020coupledTucker_hyperspectralSRR_TSP}, the multiresolution analysis-based GLPHS algorithm~\cite{aiazzi2006GLP_HS}, and the unsupervised deep learning based algorithm PAR~\cite{wei2020unsupervised}. We also considered approaches accounting for inter-image variability, including FuVar~\cite{Borsoi_2018_Fusion}, GSFus~\cite{fu2021fusionInterImageChangesL21norm} and CB-STAR~\cite{borsoi2020tensorHSRvariability}.
In this study, three real data sets with moderate variability, namely, the Ivanpah Playa, the Lake Isabella and the Lookwood, and {three} real data sets with significant variability, namely, the Lake Tahoe A and B, {and the Kern River,} were used to evaluate the performance of each method. These data sets contained one reference HRI and an MI acquired by the AVIRIS and the Sentinel-2A instruments, respectively, with a pixel of 20m resolution~\cite{Borsoi_2018_Fusion}. The HI and MI contain $L_h = 173$ and $L_m = 10$ bands, respectively.  {To illustrate the existence of the inter-image variability in the considered datasets, we computed the average absolute difference images $\frac{1}{L_m} \sum_{\ell=1}^{L_m}|\bY_{\!m}(\ell,:) - \bR(\ell,:) \bZ_h|$ (where the modulus operation $|\cdot|$ is applied elementwise), and displayed them in Figures~\ref{fig:HSIs_MSIs_a} and~\ref{fig:HSIs_MSIs_b}.}

For all acquired HRIs, which have the same spatial resolution as the MIs, a pre-processing procedure as described in~\cite{simoes2015HySure} was performed. Specifically, spectral bands that were overly noisy or corresponded to water absorption spectral regions were removed manually, and then all bands of HRIs and MIs were normalized such that the 0.999 intensity quantile corresponded to the value of 1. Moreover, all HRIs were denoised using the approach described in~\cite{roger1996denoisingHSI} to obtain a noiseless reference image $\bZ_h$. The observed HIs were generated according to~\eqref{eq:observation_model}, where $\bF$ was an $8\times 8$ Gaussian blurring operator with standard deviation~$4$ 
and $\bD$ a downsampling operator with the scaling factor~$4$. The SRF $\bR$ was acquired from calibration measurements of the Sentinel-2A instrument and known a priori. For all experiments, Gaussian noise was added to both HIs and MIs to obtain a signal-to-noise ratio (SNR) of 35~dB. To set up all baselines, we used the code provided by the authors and tuned all parameters to achieve the best fusion performance. 

We implemented the proposed DIFIV method with the CNN-based denoising engine using the PyTorch framework. The dimension of subspace $l_h$ was set to~$5$  and the number of network layers was set to~$8$, the first and hidden layers contained $l_h\times 4$ S-Conv operators while the last layer was composed by $l_h$ S-Conv operators. The Adam optimizer~\cite{kingma2014adam} with an initial learning rate $0.0002$ was used to minimize the loss function in~\eqref{eq:loss}. The number of iterations of DIFIV (Algorithm~\ref{alg}) was set to~$20$ which was sufficient to ensure convergence. The weights were initialized with the method in~\cite{he2015delving}, trained for 10000 epochs in the first iteration, and fine-tuned for 2000 epochs in the remaining iterations. We set $p = 1.5, \lambda_p = 0.01$ and $\lambda_m = \lambda_n =  0.1$ for the data with moderate variability. For the data with significant variability, we set $p = 1.8, \lambda_p = 0.002$ and $\lambda_m = \lambda_n =  0.01$. For the other parameters, we set $\rho = 0.1$ and $\epsilon = 10^{-6}$. 
\cmag{Note that in the following, the performance of the methods is compared via ${\bZ}_{h}$ only since the HRI corresponding to $\bY_{\!m}$ was not available in the experiments.}

\subsection{Quality measure and visual assessment} \label{sec:metrics}
Four quality metrics were considered to evaluate the quality of the fusion result ${\bZhat}_{h}$ compared to the ground truth ${\bZ}_{h}$. The first one is the peak signal to noise ratio (PSNR):
\begin{align}
	\text{PSNR}
	{}={} \frac{1}{L_h} \sum_{\ell=1}^{L_h}
    10\log_{10} \Bigg(
    \frac{M\, \max({{\bZ}_{h}(\ell, :)})^2}{\big\|{{\bZhat}_{h}(\ell, :) -{\bZ}_{h}(\ell, :)\big\|^2}} \Bigg)
    \nonumber\,,
\end{align}
where ${{\bZ}_{h}(\ell, :)}$ and ${\bZhat}_{h}(\ell, :)$ represent the $\ell$-th channel of ${\bZ}_{h}$ and ${\bZhat}_{h}$, respectively. 

The second metric is the Spectral Angle Mapper (SAM):
\begin{align}
	\text{SAM}
	{}={} \frac{1}{M} \sum_{m=1}^{M} \arccos \Bigg( \frac{{{\bZhat}^\top_{h}(:, m)}{{\bZ}_{h}(:, m)}}
    {\big\|{{\bZhat}_{h}(:, m)}\big\| \big\|{{\bZ}_{h}(:, m)}\big\|} \Bigg)
    \nonumber\,,
\end{align}
where ${{\bZ}_{h}(:, m)}$ and ${\bZhat}_{h}(:, m)$ denote the $m$-th pixel of ${\bZ}_{h}$ and ${\bZhat}_{h}$, respectively.

The third metric is the ERGAS~\cite{wald2000qualityERGAS}, which provides a global statistical measure of the fused image quality, defined as:
\begin{align}
\text{ERGAS}
	{}={} \frac{M}{N} \sqrt{\!\frac{10^4}{L_h} \!
    \sum_{\ell=1}^{L_h}  \frac{\big\|{\bZhat}_{h}(\ell, :) -{\bZ}_{h}(\ell, :)\big\|^2}{\mathrm{mean}({{\bZhat}_{h}(\ell, :)})^2}} 
    \nonumber \,.
\end{align}
This metric is the average of the UIQI~\cite{wang2002qualityUIQI} across bands. It evaluates image distortions including correlation loss and luminance and contrast distortions, and tends to $1$ as ${\bZhat}_{h}$ tends to ${\bZ}_{h}$.

For the visual assessment of the reconstructed images, we displayed color images at the visual spectrum (with band image at the wavelength $0.66$, $0.56$ and $0.45~\mu m$ as red, green and blue channels) and false color images at the infrared spectrum (with band image at the wavelength $2.20$, $1.50$ and $0.80~\mu m$ as red, green and blue channels). 
Due to space limitations, in the following, we only display the results of the five methods with the best quantitative performances, namely, CNMF, FuVar, GSFus, CB-STAR and DIFIV. Note that the last four algorithms account for inter-image variability.

\begin{figure*}[tp] \footnotesize
	\centering
	 CNMF \qquad \qquad \quad\ \ \ GSFus \qquad \qquad \quad \ \ CB-STAR  \qquad  \qquad \quad \ \ FuVar  \qquad \qquad \quad \ \ \ \ \ DIFIV \qquad \qquad  \quad \ \! Reference  \\
	\vspace{0.5mm}
	\includegraphics[scale=1.2]{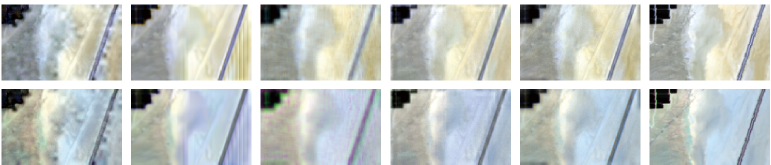}
	\caption{Visible (top) and infrared (bottom) representation for the estimated and true versions of the Ivanpah Playa HI.}
	\label{fig:playa}
\end{figure*} 
\begin{figure*}[tp] \footnotesize
	\centering
	 CNMF \qquad \qquad \quad\ \ \ GSFus \qquad \qquad \quad \ \ CB-STAR  \qquad  \qquad \quad \ \ FuVar  \qquad \qquad \quad \ \ \ \ \ DIFIV \qquad \qquad  \quad \ \! Reference  \\
	\vspace{0.5mm}
	\includegraphics[scale=1.2]{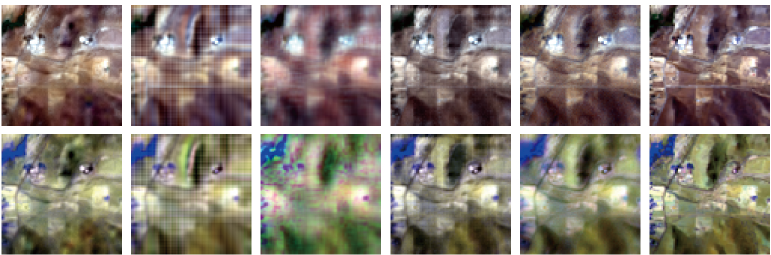}
	\caption{Visible (top) and infrared (bottom) representation for the estimated and true versions of the Lake Isabella HI.}
	\label{fig:isabella}
\end{figure*}
\begin{figure*}[tp] \footnotesize
	\centering
	 CNMF \qquad \qquad \quad\ \ \ GSFus \qquad \qquad \quad \ \ CB-STAR  \qquad  \qquad \quad \ \ FuVar  \qquad \qquad \quad \ \ \ \ \ DIFIV \qquad \qquad  \quad \ \! Reference  \\
	\vspace{0.5mm}
	\includegraphics[scale=1.2]{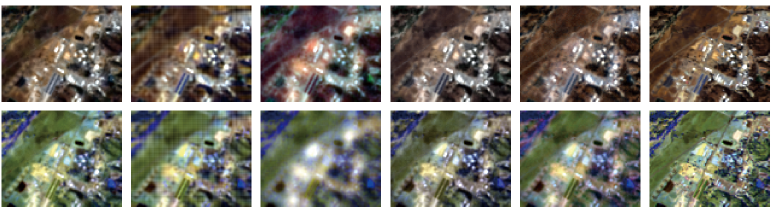}
	\caption{Visible (top) and infrared (bottom) representation for the estimated and true versions of the Lockwood HI.}
	\label{fig:lockwood}
\end{figure*}

\subsection{Category 1: Moderate variability} \label{sec:small}
In this category, we evaluated the methods using HI and MI pairs with moderate variability, including Ivanpah Playa, Lake Isabella and Lookwood. 

The first image pair considered in this category was acquired over the area surrounding Ivanpah Playa with a resolution of $80\times 128$ pixels. The second pair of images, with $80\times 80$ pixels, was captured over the Lake Isabella region, while the third pair of images containing $80\times 100$ pixels was acquired near Lookwood. The visualizations of these three image pairs {and their inter-image variability} are shown in Figure~\ref{fig:HSIs_MSIs_a}. In this category, the HI and MI look visually similar, which is typical when small differences between acquisition dates are considered (which is the case for the Lake Isabella and Lockwood images). Nevertheless, slight variations still exist, as can be seen in the overall color hue {of the} Ivanpah Playa and Lockwood images, and in the up part of the Lake Isabella {image}. 
\begin{table}[ht!]	
\caption{Results - Ivanpah Playa} \vspace{-5pt}
\centering	
\renewcommand{\arraystretch}{1.15}
{
\begin{filecontents*}{tables/ex1_playa.txt}
HySure	2.262	2.639	21.923	0.511
CNMF	1.532	2.258	23.729	0.730
GLPHS	2.924	3.139	20.949	0.508
STEREO	29.173	1643.756	17.744	0.490
SCOTT	41.025	618.314	9.388	0.307
PAR	3.506	2.260	24.011	0.752
FuVar	1.469	1.804	25.622	0.868
GSFus	1.720	1.497	27.264	0.874
CB-STAR	1.910	1.517	27.506	0.875
DIFIV	1.358	1.335	28.283	0.903
\end{filecontents*}
\pgfplotstabletypeset[header=false,
col sep = tab,
	columns/0/.style={string type},
every head row/.style={before row={ \hline}, after row=\hline},
every last row/.style={after row=\hline},
columns/0/.style={string type, column name={Algorithm}, column type/.add={@{}@{\,}}{}},
columns/1/.style={column name={SAM}, fixed, precision=3, column type/.add={@{\,}|@{\,}}{}},
columns/2/.style={column name={ERGAS}, fixed, precision=3, column type/.add={@{\,}|@{\,}}{}},
columns/3/.style={column name={PSNR}, fixed, precision=3, column type/.add={@{\,}|@{\,}}{}},
columns/4/.style={column name={UIQI}, fixed, precision=3, column type/.add={@{\,}|@{\,}}{}},
columns ={0,1,2,3,4},
every row 9 column 1/.style={postproc cell content/.style={ @cell content/.add={$\bf}{$} }},
every row 9 column 2/.style={postproc cell content/.style={ @cell content/.add={$\bf}{$} }},
every row 9 column 3/.style={postproc cell content/.style={ @cell content/.add={$\bf}{$} }},
every row 9 column 4/.style={postproc cell content/.style={ @cell content/.add={$\bf}{$} }},
]{tables/ex1_playa.txt}}
\label{tab:playa}
\end{table}

\begin{table}[ht!]	
\caption{Results - Lake Isabella} \vspace{-5pt}
\centering	
\renewcommand{\arraystretch}{1.15}
{
\begin{filecontents*}{tables/ex5_isabellalake.txt}
HySure	3.021	5.363	19.905	0.637
CNMF	2.206	3.414	25.611	0.792
GLPHS	2.755	3.572	25.207	0.793
STEREO	27.859	2145.707	19.221	0.573
SCOTT	26.281	282.097	8.453	0.076
PAR	7.482	4.044	25.454	0.805
FuVar	2.487	3.234	27.213	0.899
GSFus	2.759	3.787	26.448	0.864
CB-STAR	3.263	3.406	26.556	0.864
DIFIV	2.114	2.323	29.186	0.923
\end{filecontents*}
\pgfplotstabletypeset[header=false,
col sep = tab,
	columns/0/.style={string type},
every head row/.style={before row={ \hline}, after row=\hline},
every last row/.style={after row=\hline},
columns/0/.style={string type, column name={Algorithm}, column type/.add={@{}@{\,}}{}},
columns/1/.style={column name={SAM}, fixed, precision=3, column type/.add={@{\,}|@{\,}}{}},
columns/2/.style={column name={ERGAS}, fixed, precision=3, column type/.add={@{\,}|@{\,}}{}},
columns/3/.style={column name={PSNR}, fixed, precision=3, column type/.add={@{\,}|@{\,}}{}},
columns/4/.style={column name={UIQI}, fixed, precision=3, column type/.add={@{\,}|@{\,}}{}},
columns ={0,1,2,3,4},
every row 9 column 1/.style={postproc cell content/.style={ @cell content/.add={$\bf}{$} }},
every row 9 column 2/.style={postproc cell content/.style={ @cell content/.add={$\bf}{$} }},
every row 9 column 3/.style={postproc cell content/.style={ @cell content/.add={$\bf}{$} }},
every row 9 column 4/.style={postproc cell content/.style={ @cell content/.add={$\bf}{$} }},
]{tables/ex5_isabellalake.txt}}
\label{tab:isabella}
\end{table}

\begin{table}[ht!]	
\caption{Results - Lockwood} \vspace{-5pt}
\centering	
\renewcommand{\arraystretch}{1.15}
{
\begin{filecontents*}{tables/ex6_lockwood.txt}
HySure	3.384	4.384	22.678	0.881
CNMF	3.243	3.349	26.469	0.857
GLPHS	3.706	3.971	24.704	0.781
STEREO	28.185	883.508	21.079	0.639
SCOTT	20.109	204.538	9.273	0.094
PAR	6.610	4.433	23.634	0.754
FuVar	3.518	3.345	26.509	0.874
GSFus	3.331	3.332	26.329	0.870
CB-STAR	4.137	3.867	25.535	0.805
DIFIV	3.394	2.934	27.307	0.885
\end{filecontents*}
\pgfplotstabletypeset[header=false,
col sep = tab,
	columns/0/.style={string type},
every head row/.style={before row={ \hline}, after row=\hline},
every last row/.style={after row=\hline},
columns/0/.style={string type, column name={Algorithm}, column type/.add={@{}@{\,}}{}},
columns/1/.style={column name={SAM}, fixed, precision=3, column type/.add={@{\,}|@{\,}}{}},
columns/2/.style={column name={ERGAS}, fixed, precision=3, column type/.add={@{\,}|@{\,}}{}},
columns/3/.style={column name={PSNR}, fixed, precision=3, column type/.add={@{\,}|@{\,}}{}},
columns/4/.style={column name={UIQI}, fixed, precision=3, column type/.add={@{\,}|@{\,}}{}},
columns ={0,1,2,3,4},
every row 1 column 1/.style={postproc cell content/.style={ @cell content/.add={$\bf}{$} }},
every row 9 column 2/.style={postproc cell content/.style={ @cell content/.add={$\bf}{$} }},
every row 9 column 3/.style={postproc cell content/.style={ @cell content/.add={$\bf}{$} }},
every row 9 column 4/.style={postproc cell content/.style={ @cell content/.add={$\bf}{$} }},
]{tables/ex6_lockwood.txt}}
\label{tab:lockwood}
\end{table}

SAM, PSNR, ERGAS and UIQI metrics for all methods are reported in Table~\ref{tab:playa} to~\ref{tab:lockwood}. As shown in Table~\ref{tab:playa} and~\ref{tab:isabella}, DIFIV outperforms all competing methods in all metrics for the Ivanpah Playa and Lake Isabella images. Moreover, it can be seen in Table~\ref{tab:lockwood} that DIFIV achieves overall best results for the Lookwood data, surpassing the other methods in all metrics except for SAM, where CNMF yields the best results for this metric. Figures~\ref{fig:playa} to~\ref{fig:lockwood} illustrate the color and false color visualization of the fusion results of several algorithms. Visually, DIFIV provides the best results in recovering details and spatial reconstructions closest to the ground truth at both the visual and infrared spectra. Specifically, CNMF and GSFus introduce artifacts and fail to recover many details while CB-STAR produces blurry effects and color aberrations. FuVar and DIFIV give similar visual effects but Fuvar shows more details that do not match the reference image. This demonstrates the efficiency of DIFIV in recovering the spatial information of the latent HRIs in this category. 


\begin{figure*}[tp] \footnotesize
	\centering
	\vspace{0.5mm}	 CNMF \qquad \qquad \quad\ \ \ GSFus \qquad \qquad \quad \ \ CB-STAR  \qquad  \qquad \quad \ \ FuVar  \qquad \qquad \quad \ \ \ \ \ DIFIV \qquad \qquad  \quad \ \! Reference  \\
		\vspace{1mm}
	\includegraphics[scale=1.2]{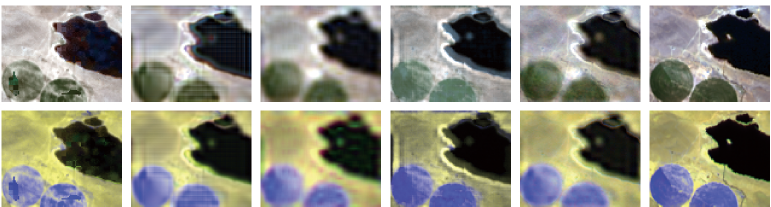}
	\caption{Visible (top) and infrared (bottom) representation for the estimated and true versions of the Lake Tahoe A HI.}
	\label{fig:tahoeA}
\end{figure*}
\begin{figure*}[tp] \footnotesize
	\centering
	 CNMF \qquad \qquad \quad\ \ \ GSFus \qquad \qquad \quad \ \ CB-STAR  \qquad  \qquad \quad \ \ FuVar  \qquad \qquad \quad \ \ \ \ \ DIFIV \qquad \qquad  \quad \ \! Reference  \\
	\vspace{0.5mm}
	\includegraphics[scale=1.2]{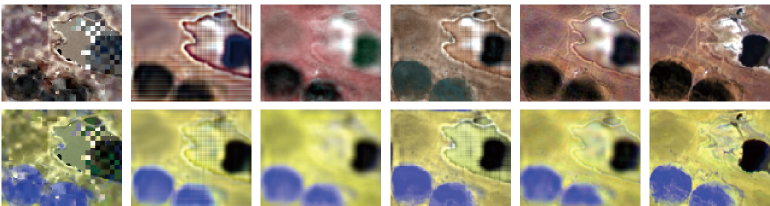}
	\caption{Visible (top) and infrared (bottom) representation for the estimated and true versions of the Lake Tahoe B HI.}
	\label{fig:tahoeB}
\end{figure*}
\begin{figure*}[tp] \footnotesize
	\centering
	 CNMF \qquad \qquad \quad\ \ \ GSFus \qquad \qquad \quad \ \ CB-STAR  \qquad  \qquad \quad \ \ FuVar  \qquad \qquad \quad \ \ \ \ \ DIFIV \qquad \qquad  \quad \ \! Reference  \\
	\vspace{1.5mm}
	\includegraphics[scale=1.2]{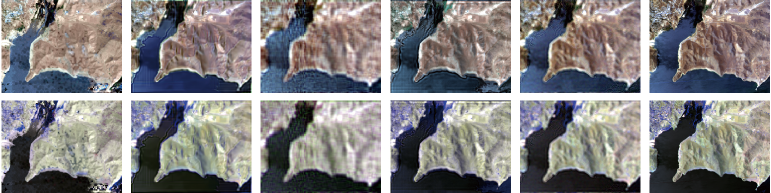}
	\caption{{Visible (top) and infrared (bottom) representation for the estimated and true versions of the Kern River HI.}}
	\label{fig:kern}
\end{figure*}
\subsection{Category 2: Significant variability} \label{sec:large}
This category evaluates the performance of the different methods when there is significant inter-image variability. We consider two image pairs acquired over the Lake Tahoe area at different time instant, namely, Lake Tahoe A and B. {Besides, an image pair captured over the Kern River scene, which comprises a larger spatial area, was also considered.}

The two {Lake Tahoe} image pairs contain $100\times 80$ pixels, {while the Kern River image pair contains $260\times 340$ pixels. The visualization of these HIs and MIs and their corresponding inter-image changes maps} can be seen in Figure~\ref{fig:HSIs_MSIs_b}. Significant variability between the HI and  MI can be easily verified in {these} cases. 
For the {two Lake Tahoe image pairs} in this category, the color hue of the ground and the crop circles is quite different. Moreover, an island on the lake is not visible in the MI of Lake Tahoe A. For Lake Tahoe B, the lake in the MI is much larger than that in the HI. {For the Kern River image pair, the river in the MI is narrower, has an upstream deposit, and shows a darker color in the water area.}

{The quantitative metrics are reported in Table~\ref{tab:tahoeA}, ~\ref{tab:tahoeB} and~\ref{tab:kern}.}  As shown in Table~\ref{tab:tahoeA}, DIFIV obtains the best results for most metrics for Lake Tahoe A and only performs slightly worse in terms of SAM compared to GSFus. It can be observed in Table~\ref{tab:tahoeB} {and~\ref{tab:kern}} that the performance of DIFIV for Lake Tahoe B and Kern River exceeds those of the competing methods for all metrics. 
A visual illustration of the fusion results for Lake Tahoe A and B in color and false color is displayed in Figure~\ref{fig:tahoeA} and Figure~\ref{fig:tahoeB}. {Figure~\ref{fig:kern} shows the visualization of the fusion results for the Kern River dataset.} It can be seen that DIFIV reconstructs more details and produces a color hue closer to the reference images at both visual and infrared spectral ranges. In particular, CNMF produced many artifacts and loses some details. GSFus and FuVar generate results with blockiness and ghosting effects while the results of CB-STAR are blurry and have some color distortions.
This demonstrates the superiority of DIFIV in recovering the latent HRIs when significant variability exists. 


\begin{table}[ht!]	
\caption{Results - Lake Tahoe A} \vspace{-5pt}
\centering	
\renewcommand{\arraystretch}{1.15}
{
\begin{filecontents*}{tables/ex4_tahoeA.txt}
HySure	10.643	7.775	16.531	0.655
CNMF	12.371	7.514	18.102	0.676
GLPHS	10.803	7.206	18.303	0.701
STEREO	30.605	2541.149	15.991	0.575
SCOTT	42.839	457.101	9.243	0.215
PAR	15.886	6.065	20.579	0.811
FuVar	8.373	6.545	19.258	0.780
GSFus	6.628	4.376	22.537	0.883
CB-STAR	7.548	3.769	24.165	0.917
DIFIV	6.737	3.706	24.174	0.922
\end{filecontents*}
\pgfplotstabletypeset[header=false,
col sep = tab,
	columns/0/.style={string type},
every head row/.style={before row={ \hline}, after row=\hline},
every last row/.style={after row=\hline},
columns/0/.style={string type, column name={Algorithm}, column type/.add={@{}@{\,}}{}},
columns/1/.style={column name={SAM}, fixed, precision=3, column type/.add={@{\,}|@{\,}}{}},
columns/2/.style={column name={ERGAS}, fixed, precision=3, column type/.add={@{\,}|@{\,}}{}},
columns/3/.style={column name={PSNR}, fixed, precision=3, column type/.add={@{\,}|@{\,}}{}},
columns/4/.style={column name={UIQI}, fixed, precision=3, column type/.add={@{\,}|@{\,}}{}},
columns ={0,1,2,3,4},
every row 7 column 1/.style={postproc cell content/.style={ @cell content/.add={$\bf}{$} }},
every row 9 column 2/.style={postproc cell content/.style={ @cell content/.add={$\bf}{$} }},
every row 9 column 3/.style={postproc cell content/.style={ @cell content/.add={$\bf}{$} }},
every row 9 column 4/.style={postproc cell content/.style={ @cell content/.add={$\bf}{$} }},
]{tables/ex4_tahoeA.txt}}
\label{tab:tahoeA}
\end{table}

\begin{table}[ht!]	
\caption{Results - Lake Tahoe B} \vspace{-5pt}
\centering	
\renewcommand{\arraystretch}{1.15}
{
\begin{filecontents*}{tables/ex4_tahoeB.txt}
HySure	13.458	12.042	11.913	0.235
CNMF	7.954	7.289	16.387	0.428
GLPHS	6.662	4.786	19.824	0.665
STEREO	29.877	7936.808	15.208	0.463
SCOTT	42.427	491.817	7.504	0.136
PAR	11.787	6.210	21.405	0.728
FuVar	4.688	3.729	21.860	0.790
GSFus	4.182	3.160	23.425	0.826
CB-STAR	3.950	2.597	25.221	0.881
DIFIV	3.265	2.396	25.834	0.899
\end{filecontents*}
\pgfplotstabletypeset[header=false,
col sep = tab,
	columns/0/.style={string type},
every head row/.style={before row={ \hline}, after row=\hline},
every last row/.style={after row=\hline},
columns/0/.style={string type, column name={Algorithm}, column type/.add={@{}@{\,}}{}},
columns/1/.style={column name={SAM}, fixed, precision=3, column type/.add={@{\,}|@{\,}}{}},
columns/2/.style={column name={ERGAS}, fixed, precision=3, column type/.add={@{\,}|@{\,}}{}},
columns/3/.style={column name={PSNR}, fixed, precision=3, column type/.add={@{\,}|@{\,}}{}},
columns/4/.style={column name={UIQI}, fixed, precision=3, column type/.add={@{\,}|@{\,}}{}},
columns ={0,1,2,3,4},
every row 9 column 1/.style={postproc cell content/.style={ @cell content/.add={$\bf}{$} }},
every row 9 column 2/.style={postproc cell content/.style={ @cell content/.add={$\bf}{$} }},
every row 9 column 3/.style={postproc cell content/.style={ @cell content/.add={$\bf}{$} }},
every row 9 column 4/.style={postproc cell content/.style={ @cell content/.add={$\bf}{$} }},
]{tables/ex4_tahoeB.txt}}
\label{tab:tahoeB}
\end{table}

\begin{table}[ht!]	
\caption{{Results - Kern River}} \vspace{-5pt}
\centering	
\renewcommand{\arraystretch}{1.15}
{
{\begin{filecontents*}{tables/ex7_kern.txt}
HySure	9.094	8.933	21.717	0.442
CNMF	5.851	8.471	22.853	0.356
GLPHS	8.231	7.279	24.190	0.492
STEREO	30.337	636.136	22.568	0.450
SCOTT	27.652	220.140	13.239	0.045
PAR	11.695	6.742	28.000	0.739
FuVar	4.654	5.144	28.335	0.797
GSFus	5.037	4.243	29.404	0.785
CB-STAR	5.298	5.004	28.884	0.729
DIFIV	3.412	3.734	31.506	0.852
\end{filecontents*}
\pgfplotstabletypeset[header=false,
col sep = tab,
	columns/0/.style={string type},
every head row/.style={before row={ \hline}, after row=\hline},
every last row/.style={after row=\hline},
columns/0/.style={string type, column name={Algorithm}, column type/.add={@{}@{\,}}{}},
columns/1/.style={column name={SAM}, fixed, precision=3, column type/.add={@{\,}|@{\,}}{}},
columns/2/.style={column name={ERGAS}, fixed, precision=3, column type/.add={@{\,}|@{\,}}{}},
columns/3/.style={column name={PSNR}, fixed, precision=3, column type/.add={@{\,}|@{\,}}{}},
columns/4/.style={column name={UIQI}, fixed, precision=3, column type/.add={@{\,}|@{\,}}{}},
columns ={0,1,2,3,4},
every row 9 column 1/.style={postproc cell content/.style={ @cell content/.add={$\bf}{$} }},
every row 9 column 2/.style={postproc cell content/.style={ @cell content/.add={$\bf}{$} }},
every row 9 column 3/.style={postproc cell content/.style={ @cell content/.add={$\bf}{$} }},
every row 9 column 4/.style={postproc cell content/.style={ @cell content/.add={$\bf}{$} }},
]{tables/ex7_kern.txt}}}
\label{tab:kern}
\end{table}

\subsection{{Parameter Sensitivity}} \label{sec:parameter}
\begin{figure}[tp]
	\centering
    \includegraphics[trim = 0mm 3mm 2mm 2mm, clip, scale=0.7]{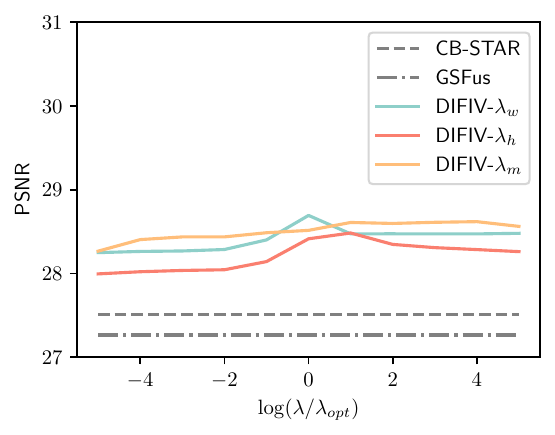}
	\vspace{-2mm}
	\caption{{Sensitivity of the proposed DIFIV method with respect to regularization parameters $\lambda_p, \lambda_h, \lambda_m$.}}
	\label{fig:para}
\end{figure}

In this subsection, we study the sensitivity of DIFIV to the choice of values for regularization parameters $\lambda_p, \lambda_h, \lambda_m$. Considering the Ivanpah Playa scene as an example, we varied each parameter individually while keeping the remaining ones fixed at the values described in Subsection~\ref{sec:sim_setup}. The PSNR values of the fusion results as a function of the ratio $\log_{10}(\lambda / \lambda_{opt})$ are shown in Figure~\ref{fig:para}, where $\lambda_{opt}$ is the empirically selected value of the corresponding parameters. The PSNR values of two selected competing methods (CB-STAR and GSFus) are also shown for reference. It can be observed that varying parameters of DIFIV even by various orders of magnitude only leads to moderate variations of PSNR values, which are consistently higher than that of the competing methods. \cred{Moreover, the parameters of GSFus and CB-STAR were adjusted to provide the best performance in each example, and their performance would likewise degrade if their parameters move away from their optimal values, as discussed in the original works~\cite{fu2021fusionInterImageChangesL21norm,borsoi2020tensorHSRvariability}.} This indicates the performance of DIFIV is not overly sensitive to the choice of regularization parameters.

\subsection{{Computational cost}} \label{sec:computation}
{This experiment aims at comparing the computational cost of the algorithms accounting for inter-image variability. DIFIV was implemented using Python while the remaining methods were implemented using MATLAB. We conducted all the experiments on a computer with an Intel Core i7-10700 CPU, 32-GB random access memory and an NVIDIA Quadro P2200 GPU. The execution times of the algorithms for all the tested image pairs are shown in Table~\ref{tab:time}. It can be seen that the computation times of DIFIV are substantially higher than those of the competing methods, which comes as a compromise for its superior image fusion quality results. Nevertheless, the computation times of DIFIV scale reasonably with the image sizes; for instance, comparing the results for the Lake Isabella and Kern River images, we see that an increase of about ten times in the number of pixels in the image leads to an increase of about two times in the computation times. The development of computationally efficient extensions to the DIFIV method will be investigated in future work.}


\begin{table}[t] \footnotesize	
\caption{Execution times of the algorithms that consider inter-image variability (in seconds)} \vspace{-5pt}
\centering	
\renewcommand{\arraystretch}{1.15}
\begin{tabular}{c|ccccccc}
		\hline
		{} & FuVar & GSFus & CB-STAR & DIFIV \\ \hline
        Ivanpah Playa & 354.6 & 31.4 & 11.1 & 2963.4 \\
        Lake Isabella & 199.3 & 17.7 & 8.8 & 2928.7 \\
        Lockwood & 228.8 & 23.2 & 30.1 & 2954.0 \\
        Lake Tahoe~A & 679.5 & 23.1 & 7.8 & 2178.4 \\
        Lake Tahoe~B & 718.9 & 22.2 & 7.6 & 2143.5 \\
        Kern River & 1762.0 & 307.3 & 96.0 & 5908.4 \\
		\hline
	\end{tabular}
\label{tab:time}
\end{table}

\section{Conclusions}
\label{sec:con}
This paper presented an unsupervised deep learning-based HMIF method accounting for inter-image variability. We first formulated a new imaging model considering both the joint as well as the image-specific priors related to the two latent HRIs. The inter-image variations were modeled using a hyper-Laplacian distribution, while the image-specific priors of the latent HRIs were defined implicitly by deep denoising engines. An iteratively reweighted scheme was then investigated to solve the non-convex cost function and tackle the joint image prior term. The optimization problem was solved using a variable splitting strategy, and the deep image priors were implemented by means of CNN-based denoising operations. A lightweight, image-specific CNN-based denoiser with a zero-shot training strategy was designed. The network parameters were iteratively updated during the optimization procedure in order to adapt to variations in the statistical properties of the estimated HRIs as the method converged. The proposed method achieved superior experimental performance in the presence of both moderate and significant inter-image variability when compared to state-of-the-art approaches.


\bibliographystyle{IEEEtran}
\bibliography{references,references_old}

\end{document}